\documentclass[12pt,dvips]{article}
\usepackage[dvips]{graphicx}
\usepackage{graphicx}
\usepackage{cite}
\usepackage{amssymb}

\graphicspath{{plot/}}
\DeclareGraphicsExtensions{.eps.gz,.eps,.ps,.ps.gz}

\def\lsim{\raise0.3ex\hbox{$<$\kern-0.75em\raise-1.1ex\hbox{$\sim$}}}
\def\gsim{\raise0.3ex\hbox{$>$\kern-0.75em\raise-1.1ex\hbox{$\sim$}}}

\parskip=\itemsep               %?
\newcommand{\lambdabar}{{\hbox{$\lambda_e$\kern-1.9ex\raise+0.45ex\hbox{--}
\kern+0.2ex}}}

\newif\ifhepph
 
\hepphtrue

\newcommand{\Enu}{E_{\overline{\nu}}}

\newcommand{\Enubar}{\epsilon_{\overline{\nu}}}

\newcommand{\Fnu}{{\cal F}_{\overline{\nu}}}
\newcommand{\tbar}{\overline{\tau}_n}

\ifhepph\date{\empty}\fi

\title{
\ifhepph{\normalsize\rightline{DESY 04-192}\rightline{NUB-3251-Th-04}
\rightline{WUB 04-12}\rightline{ITP-BUDAPEST 615}}\fi
\vskip 1cm
\bf\boldmath
Upper Bounds on the Neutrino-Nucleon Inelastic Cross Section 
       \vspace{21mm}}
\author{
L.~A.~Anchordoqui$^{a,b}$,
Z.~Fodor$^{c,d}$, S.~D.~Katz$^c$\thanks{On leave from Institute for Theoretical Physics, E\"otv\"os University,
Budapest, Hungary.}, A.~Ringwald$^a$, and H.~Tu$^a$\\[1cm]
\it $^a$Deutsches Elektronen-Synchrotron DESY,
Hamburg, Germany\\
\it $^b$Department of Physics, Northeastern University, Boston MA 02155, USA\\
\it $^c$Department of Physics, University of Wuppertal,
Germany\\
\it $^d$Institute for Theoretical Physics, E\"otv\"os University,
Budapest, Hungary
}
 
\begin{document}
\begin{titlepage}
  \maketitle

\begin{abstract}

\noindent
Extraterrestrial neutrinos can initiate deeply developing air showers, 
and those that traverse the atmosphere unscathed may produce 
cascades in the ice or water. Up to now, no such events have been observed.
This can be translated into upper limits on the diffuse neutrino flux. 
On the other hand, the observation of cosmic rays with primary energies~$> 10^{10}$~GeV 
suggests that there is a guaranteed flux of cosmogenic 
neutrinos, arising from the decay of charged pions (and their muon daughters)  
produced in proton interactions with the cosmic microwave background. In 
this work, armed with these cosmogenic neutrinos and the increased 
exposure of neutrino telescopes we bring up-to-date model-independent upper 
bounds on the neutrino-nucleon inelastic cross section. Uncertainties in 
the cosmogenic neutrino flux are discussed and taken into account in our 
analysis. The prospects for improving these bounds with the Pierre Auger 
Observatory are also estimated. The unprecedented statistics to be collected 
by this experiment in 6~yr of operation will probe the neutrino-nucleon inelastic cross 
section at the level of Standard Model predictions.

\end{abstract}
 
\thispagestyle{empty}
\end{titlepage}
\newpage \setcounter{page}{2}
 
\section{\label{intro}Introduction}

Ultrahigh energy cosmic neutrinos may reveal aspects of nature hidden to us so far.
They can point back to very distant sources, resolving the origin of the highest
energy cosmic rays and the underlying acceleration mechanism.
Upon their arrival at Earth, they interact with nucleons at centre-of-mass energies 
around several hundreds of TeV, probing the energy regime far beyond the reach of
terrestrial colliders.

The detection of ultrahigh energy cosmic neutrinos is a very challenging task
due to their weak interactions.
Although the neutrino-nucleon inelastic  cross section increases with energy, 
the cosmic neutrino flux falls even more steeply with energy.
Large-scale projects and novel techniques are being deployed and proposed.
The Pierre Auger Observatory (PAO)~\cite{Abraham}, 
followed by IceCube~\cite{Ahrens:2002dv}, and possibly 
EUSO~\cite{Catalano:2001mm} and 
OWL~\cite{Stecker:2004wt}, will 
reach a sensitivity of the level of theoretical predictions for the cosmic 
neutrino fluxes. Moreover, before the next generation experiments lead us 
into the exciting era,
the non-observation of neutrino-induced events reported by the 
Fly's Eye~\cite{Baltrusaitis:mt},
the AGASA~\cite{Inoue:cn,Yoshida:2001icrc} and the 
RICE~\cite{Kravchenko:2002mm,Kravchenko:2003tc} collaborations can be used to
set upper limits on the diffuse neutrino flux.
Additionally, by exploiting a certain prediction for the neutrino flux, the 
search results can also be 
turned into upper bounds on the neutrino-nucleon inelastic cross 
section~\cite{Berezinsky:1974kz,Anchordoqui:2002vb,Tu:2004ms}.

Ultrahigh energy cosmic neutrinos from diverse sources are predicted, 
and their existence is strongly supported by the observation of the 
ultrahigh energy cosmic rays (UHECRs).
Among all, the so-called cosmogenic neutrinos \cite{Berezinsky:1969}
are almost guaranteed to exist. 
They originate from the decay of charged pions produced in the interactions 
of protons with the cosmic microwave background (CMB).
There are still  some uncertainties in the estimate of the cosmogenic 
neutrino flux.  This is due to our poor knowledge of the nature and the 
origin of the UHECRs. Possible ranges for the size of the cosmogenic 
neutrino flux have been investigated elsewhere~\cite{Fodor:2003ph}   
assuming that the cosmic ray spectrum beyond $10^{8}$~GeV is dominated by 
extragalactic protons with an isotropic distribution of sources.
In this work, armed with these cosmogenic neutrinos and  
the increased exposure of neutrino detectors, we show that the 
neutrino-nucleon inelastic cross section is tighter constrained than 
previously noted. Uncertainties in the cosmogenic neutrino flux are discussed 
and taken into account in our analysis.
The prospects for improving these bounds with the PAO 
are also estimated.

The structure of the paper is the following. In Sec.~\ref{formulae} we examine acceptances 
for neutrino detection. In particular, we compute the effective apertures of AGASA and
RICE as examples of ground arrays and under-ice neutrino telescopes,
respectively. We also estimate the aperture of the PAO in order to investigate future sensitivity
to physics beyond the Standard Model (SM). In Sec.~\ref{flux} we present an overview of cosmogenic neutrino 
fluxes. In Sec.~\ref{table} we derive 
model-independent upper bounds on the neutrino-nucleon
inelastic cross section from the search results reported by the AGASA
and the RICE collaborations.
In Sec.~\ref{PAO}, the prospects for improving these bounds with the PAO are
estimated.
Section~\ref{conclusions} comprises our conclusions.

\section{\label{formulae} Search for Ultrahigh Energy Cosmic Neutrinos}

Due to their feeble interaction and flux, ultrahigh energy cosmic neutrinos 
are extremely difficult to detect.
To overcome this difficulty one either needs a huge target volume for the detectors, 
or one has to employ novel detection techniques.
Besides, the unwanted strong backgrounds from cosmic ray protons must be effectively 
reduced.
Following these considerations, 
ultrahigh energy cosmic neutrinos are searched for in the Earth 
atmosphere\cite{Baltrusaitis:mt,Inoue:cn,Yoshida:2001icrc}, 
in the Greenland~\cite{Lehtinen:2003xv} and Antarctic~\cite{Ahrens:2003ee} ice sheet, in the sea/lake~\cite{Balkanov:2001mk}, or even in the regolith 
of the moon~\cite{Gorham:2003da}. For this purpose fluorescence detectors, 
ground arrays, underwater 
and -ice neutrino telescopes exist in various stages of maturity, from proposed to 
nearly completed.
In this section we summarise general formulae for estimating neutrino-induced event rates.
We discuss separately their applicability in the case when neutrinos interact significantly 
more strongly than in the SM, as predicted in many 
new physics scenarios~\cite{Domokos:1998ry}.

\begin{figure}[t!]
\vspace{0.6cm}
\begin{center}
\includegraphics*[width=9.5cm,clip=]{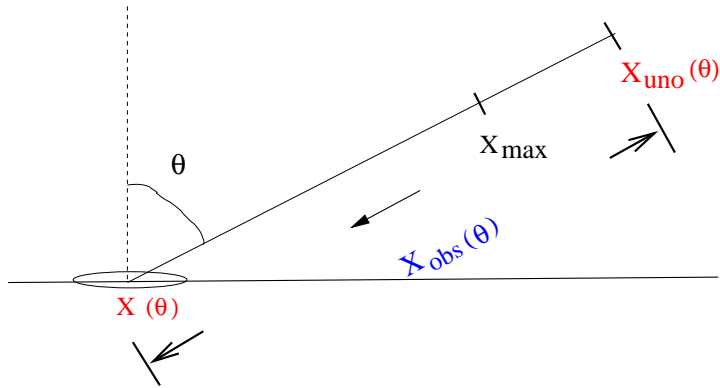}
%\vspace{-1.0cm}
\caption[dum]
{Schematic diagram for our definition for the
range of depths $X_{\rm obs} (\theta)$
for a ground array situated at an atmospheric slant depth $X (\theta)$.
Showers initiated below $X_{\rm uno} (\theta)$ are visible to the ground array.
On average, a $10^{10}$~GeV shower develops to its maximum 
$X_{\rm max}$ after traversing about $800~{\rm g / cm^2}$ 
in the atmosphere.
}
\label{range_of_depths}
\end{center}
\end{figure}

We start with 
the differential rate of air showers initiated at the point $(\ell,\theta)$
by particles (cosmic ray protons, cosmic neutrinos etc.) incoming with energy $E$ and
of flux $F (E)$,
where $\ell$ is the distance of this point to the detector measured
along the shower axis, and $\theta$ is the angle to the zenith at the point where the shower
axis hits the Earth's surface (cf. Fig.~\ref{range_of_depths}).
The number of air showers induced due to the interaction, the 
cross section for which is $\sigma (E)$,
per unit of time $t$, area $A$,
solid angle $\Omega$ (with $d \Omega = \sin \theta\, d \theta\, d \phi$) and
energy $E_{\rm sh}$ is
\begin{equation}
\label{eq:shower_prob}
   \frac{d}{d \ell} \left(
   \frac{d^4 N}{dt\, dA\, d\Omega\, dE_{\rm sh}} \right)
   = \frac{1}{m_p}\, \sigma (E)\, F (E)\,
   e^{ - \frac{\sigma^{\rm tot} (E)\,
   x (\ell, \theta)}{m_p}}\, \rho_{\rm air} [h(\ell, \theta)]\, ,
\end{equation}
where $m_p$ is the proton mass, and $\rho_{\rm air}$ is the air density at the
altitude $h$.
The relation of the energy deposited in a visible shower, $E_{\rm sh}$, 
to the incident particle energy $E$ is dependent on the scattering process. 
For protons, $E_{\rm sh} \approx E$, whereas in general, $E_{\rm sh} = y  E$, where 
$y$ is the inelasticity. For neutrino interactions, the inelasticity distribution is measured
through deep inelastic scattering processes for which $y = (E_{\nu} - E^\prime_l) / E_{\nu},$
where $E^\prime_l$ is the energy of the final state lepton. For simplicity, throughout 
this paper we take the average value of this distribution, denoted by $\left< y \right>$.
The exponential term in Eq.~(\ref{eq:shower_prob})
accounts for the flux attenuation in the atmosphere, where
the total inelastic cross section 
$\sigma^{\rm tot} (E) = \sigma^{\rm SM} (E) + \sigma^{\rm new} (E)$ can receive
contributions from SM and new physics interactions.\footnote{In general, the inelasticities of the
Standard Model and of the new physics contributions might be different.
For $\sigma^{\rm new} \gg \sigma^{\rm SM}$, as considered in our analysis, one is sensitive only to
the $\left< y \right>$ of the new physics.}

After carrying out the integral of 
$\rho_{\rm air} (\ell, \theta)\, d \ell \equiv -\, d x (\ell, \theta)$ 
over the range of depths, $X_{\rm obs} (\theta) \equiv X (\theta) - X_{\rm uno} (\theta)$, 
within which showers induced are visible to the ground array 
detectors, the rate of neutrino-induced events at a ground array with threshold energy 
$E_{\rm th}$ can be well 
estimated as
\cite{Tu:2004ms}
\begin{eqnarray}
\label{eq:AGASA_check}
\hspace{-2cm}
   N &=& t \int_{E_{\rm th}}  d E_{\rm sh}\, 
   \frac{\sigma_{\nu N} (E_{\nu})}{\sigma^{\rm tot}_{\nu N} (E_{\nu})}\, 
   F_{\nu} (E_{\nu})\nonumber \\ 
   & \times &  \int^{\theta_{\rm max}}_{\theta_{\rm min}} d \theta\,
    {\cal S} (E_{\rm sh}, \theta)\, 2 \pi  %\cos \theta %270804ht
   \sin \theta\, \left(e^{- \frac{X_{\rm uno} (\theta)\,
   \sigma^{\rm tot}_{\nu N} (E_{\nu})}{m_p}} - e^{- \frac{X (\theta)\,
   \sigma^{\rm tot}_{\nu N} (E_{\nu})}{m_p}} \right) \nonumber \\
   &\approx& t\, A_p\, \int_{E_{\rm th}}\, d E_{\rm sh}\, 
   \frac{\sigma_{\nu N} (E_{\nu})}{\sigma^{\rm tot}_{\nu N} (E_{\nu})}\,
   F_{\nu} (E_{\nu})\, P (E_{\rm sh})\, {\rm att} (E_{\nu})\, , 
\end{eqnarray}
where 
$X (\theta)$ is the atmospheric slant depth of the ground array, 
$X_{\rm uno} (\theta)$ is the minimum atmospheric depth a neutrino must reach in order
to induce an observable shower to the ground array
(cf. Fig.~\ref{range_of_depths}), and
\begin{equation}
 {\rm att} (E_{\nu})   \equiv  \int^{\cos \theta_{\rm min}}_{\cos \theta_{\rm max}}
   d \cos \theta\, 2 \pi\, \cos \theta\, e^{- \frac{X_{\rm uno}
   (\theta)\, \sigma^{\rm tot}_{\nu N} (E_{\nu})}{m_p}}\,
   \left(1 - e^{- \frac{X_{\rm obs} (\theta)\,
   \sigma^{\rm tot}_{\nu N} (E_{\nu})}{m_p}} \right) \,\,.
\end{equation}
We approximated the effective area as ${\cal S} (E, \theta) \approx
A_p\, P (E)\, \cos \theta$,  
with $A_p$ a parameter which is of the same order of the detector's 
geometric area.

In the SM, the hadronic component produced in the charged current
($\nu N \rightarrow l X$) and neutral current neutrino-nucleon interactions 
($\nu N \rightarrow \nu X$) generates a shower which is visible at
the ground array. 
The hadronic component inherits an energy $E_{\rm sh} = \left< y \right> E_{\nu}$ from the neutrino. 
 The charged current process 
for electron neutrinos produces both a hadronic shower ($E_{\rm sh} \approx 0.2 E_\nu$)
and an electromagnetic shower ($E_{\rm sh} \approx 0.8 E_\nu$). Thus, in such a case all the energy is 
deposit in the atmosphere. 

Note that the neutrino-nucleon inelastic cross section 
$\sigma_{\nu N} (E_{\nu})$ is also contained in the
(over zenith angle integrated) ``attenuation factor'' ${\rm att} (E_{\nu})$. 
This factor determines the effective aperture for neutrino detection.
In the SM, neutrinos interact weakly 
(e.g. $\sigma^{CC}_{\nu N} \approx 10^{-4}~{\rm mb}$ 
at $E_{\nu} = 10^{11}$~GeV), 
so Eq.~(\ref{eq:AGASA_check}) can be reduced to the more familiar formula
\begin{equation}
   N \approx \frac{t}{m_p} \int d E_{\rm sh}\, d \Omega\, 
   \sigma_{\nu N} (E_{\nu})\, F_{\nu} (E_{\nu})\,
   {\cal S} (E_{\rm sh}, \theta)\, X_{\rm obs} (\theta)\, .   
\label{N4}
\end{equation}
It is obvious 
that ground arrays should look at the quasi-horizontal direction,
i.e. $\theta \gtrsim 70^{\circ}$, in order to achieve a large range of observability
$X_{\rm obs}$, and thus a
larger neutrino detection rate. At the same time,
it helps to reduce the backgrounds from cosmic ray protons.

For an under-ice neutrino telescope, the hadronic and electromagnetic cascades are 
detectable only when they are initiated within or very close to the detector.
The rate for the contained events per solid angle $\Omega$ is
\begin{equation}
\label{rate_rice}
   \frac{d^2 N}{d t\, d \Omega} = 
   \frac{\rho_{\rm ice}}{m_p} \int_{E_{\rm th}} d E_{\rm sh}\, 
   F_{\nu} (E_{\nu})\, \sigma_{\nu N} (E_{\nu})\, V (E_{\rm sh})\, 
   e^{ - \frac{\sigma^{\rm tot}_{\nu N} (E_{\nu})\, x (\theta)}{m_p}}\, ,
\end{equation}
where $\rho_{\rm ice} = 0.92~{\rm g / cm^3}$ is the ice density, and 
$V$ is the effective volume of the detector, which is usually energy
dependent.  
The exponential function takes into account the neutrino flux attenuation 
in the atmosphere and in the ice. Here,
$x (\theta) = X_{\rm atm} (\theta) + \rho_{\rm ice}\, d (\theta)$, where
$d$ is the distance traversed
by neutrinos incident with zenith angle $\theta$ (measured from the centre of the detector)
from the ice upper surface to the detector's surface.

The flux attenuation in the ice is much more effective than in the atmosphere. 
The number of neutrinos reaching the detector's target volume will be 
strongly reduced if the neutrino-nucleon cross section is enhanced.
On the other hand, at sufficiently high energies, under-ice neutrino telescopes 
are also sensitive to 
interaction vertices located several kilometers away which produces muon(s) or tau(s) 
passing through the detector (through-going muon and tau events). 
This increases their effective aperture for neutrinos largely.  
However, beyond the SM, the predictions of the muon or tau spectra depend strongly on the   
scenario considered (see e.g. \cite{Kowalski:2002gb}).

%%%%%%%%%%%%%%%%%%%%%%%%%%%%%%%%FIGURE%%%%%%%%%%%%%%%%%%%%%%%%%%
%%%%%%%%%%%%%%%%%%%%%%%%%%%%%%%%%%%%%%%%%%%%%%%%%%%%%%%%%%%%%%%%%

Tau neutrinos can also be searched for through the showers
induced by tau lepton decay in the atmosphere \cite{Fargion:2000iz,Feng:2001ue}.
Though conventional astrophysical sources do not produce tau neutrinos, 
terrestrial experiments (see
e.g.~\cite{Fukuda:2000np}) have shown that $\nu_\mu$ and $\nu_\tau$
are maximally mixed with a mass-squared difference $\sim
10^{-3}~{\rm eV}^2$. This, together with the known smallness
of $|\langle \nu_e|\nu_3\rangle|^2$~\cite{Bilenky:2001jq},
implies that the $\nu_\mu$'s will partition themselves equally
between $\nu_\mu$'s and $\nu_\tau$'s on lengths
large compared to the oscillation length $\lambda_{\rm osc}\sim
1.5 \times 10^{-3}\,({E_\nu}/{\rm PeV})$~pc;
here $\nu_3 =(\nu_\mu+\nu_\tau)/\sqrt{2}$ is the third neutrino eigenstate.
One consequence of this remarkable symmetry is the process of decohereing: 
any initial flavour ratio $w_\alpha$ ($\alpha = e, \mu, \tau$) that contains 
$w_e = 1/3$ will arrive at Earth with equipartition on the three flavours~\cite{Learned:1994wg}. 
Since cosmic neutrinos originate dominantly from the decay of $\pi^\pm$ 
and their muon daughters, their initial flavour ratios 
$w_e : w_\mu : w_\tau \equiv 1 : 2 : 0,$ 
should arrive at Earth democratically distributed, i.e., $1: 1: 1.$

If the geometrical configuration is met, tau neutrinos entering the Earth at large nadir 
angles will convert to a tau lepton near Earth's surface.
The tau lepton exits the Earth and then decays in the surroundings of the array.  
The detection rate depends on the neutrino survival probability after a 
distance $z$ in the Earth for conversion to a 
charged lepton, and 
on the probability the charged lepton would escape the Earth's surface with 
an energy above the detector's 
threshold, i.e.~\cite{Feng:2001ue}
\begin{equation}
   \frac{d^2 N}{d t\, d E_{\tau}} \propto \int d z\,\,
e^{- \int^{z}_0 d z^\prime\, \frac{\sigma^{\rm tot}_{\nu N} (E_{\nu})\, \rho[r (\theta, z^\prime)]}{m_p}}\, \,\,
\frac{\sigma^{CC}_{\nu N} (E_{\nu})\, \rho_s}{m_p}\,
    P_{\rm esc} (E_{\tau})\,  
   \cdot (A \Omega)_{\rm eff} (E_{\tau})\, , 
\label{N6}
\end{equation}
where $\rho (r)$ is Earth density at radius $r$, and 
$\rho_s \approx 2.65~{\rm g / cm^3}$ Earth's surface density. 
The effective aperture $(A \Omega)_{\rm eff}$ depends on the probability of
the $\tau$ inducing a shower which triggers the detector.

While improving the prospects of neutrino detection to a large extent, 
this signal is not always present 
when neutrino interactions predicted by scenarios beyond the SM 
are considered. 
An enhancement of the neutrino-nucleon total cross section will lead to a decrease of the
neutrino interaction length in the Earth. 
This will make the geometrical configuration even more difficult to meet.
Though the neutrino interaction probability is enhanced, 
the hadronic interaction products will be absorbed in the Earth. 
The leptons which can escape the Earth may only
inherit a tiny fraction of the incident neutrino energy, so that the shower they induce is 
practically undetectable.
This has been demonstrated in~\cite{Anchordoqui:2001cg} with the example of  
black hole production in ultrahigh energy neutrino-nucleon scattering.
Nevertheless, the earth-skimming tau neutrinos provide a way to discriminate 
new physics interactions from SM ``backgrounds''.

Up to now, experiments with ultrahigh energy neutrino sensitivity include 
AGASA~\cite{Inoue:cn,Yoshida:2001icrc},
RICE~\cite{Kravchenko:2002mm,Kravchenko:2003tc}, FORTE~\cite{Lehtinen:2003xv}, 
GLUE~\cite{Gorham:2003da},  Fly's Eye~\cite{Baltrusaitis:mt}, 
and AMANDA~\cite{Ahrens:2003ee}. In the following we consider  AGASA and
RICE as 
case studies: {\em i)} AGASA is
sensitive to deeply developing showers in the atmosphere and uses a well
developed technique; {\em ii)}
RICE has a greater exposure, though the experimental technique has a
shorter track record.  In the future,
the best measurements will come from the PAO~\cite{Capelle:1998zz} 
and IceCube~\cite{Ahrens:2003ix} experiments,
which have a similar aperture
at these energies;  below we consider the PAO as an example of future
sensitivity to physics beyond the SM.

\subsection{AGASA}

The Akeno Giant Air Shower Array (AGASA) occupies farm land near the village of Akeno (Japan) 
at a longitude of $138^\circ 30'$ East and  latitude $35^\circ 30'$ 
North~\cite{Chiba:1991nf}. 
The array, which consists of 111 
surface detectors deployed over an area of about 100 km$^2$, has been running since 1990. 
About 95\% of the surface detectors were operational from March to December 
1991, and
the array has been fully operational since then.  A prototype detector operated from 
1984 to 1990 and has been part of AGASA since 1990~\cite{Teshima:1985vs}.

The AGASA Collaboration has searched for deeply-penetrating  inclined 
(zenith angle $\theta > 60^{\circ}$) air showers.
Non-observation of neutrino events during a running time of 
$9.7 \times 10^7$~s was reported 
in~\cite{Inoue:cn}, implying,  for zero events of background,  
an upper limit of 2.44 at 90$\%$ confidence limit (CL)~\cite{Feldman:1997qc}.
The AGASA data for deeply penetrating air showers recorded from December 1995
to November 2000 (with an effective lifetime of
1710.5 days) 
is published in~\cite{Yoshida:2001icrc}.
Deeply penetrating events must satisfy following search criteria:
{\em i)} $\theta \geq 60^{\circ}$,
{\em ii)} $|X^{\eta}_{\rm max} - X_{\rm slant}| \leq 500~{\rm g} / {\rm cm}^2$,
where $X_{\rm slant}$ is the atmospheric slant depth of the AGASA array centre along
the shower axis, and {\em iii)}
$X^{\eta}_{\rm max},\,\, X^{\delta}_{\rm max} \geq 2500~{\rm g} / {\rm cm}^2$,
where $\eta$ and $\delta$ parametrise the lateral distribution of shower particle
densities and the curvature of the shower disc front, respectively.
By fitting them to the empirical formulae, the shower maximum $X_{\rm max}$ can possibly
be deduced.
Specifically, there was one event observed, consistent with
the expected background from hadronic showers
1.72$^{+0.14+0.65}_{-0.07-0.41}$ (MC statistics and systematic).
The AGASA search result therefore corresponds to an upper bound of 3.5 events at 
95$\%$ CL~\cite{Feldman:1997qc}.
In Ref.~\cite{Anchordoqui:2002vb}, 
the AGASA search result has been combined with the non-observation of deeply-penetrating
showers reported by the Fly's Eye Collaboration \cite{Baltrusaitis:mt}
to set model independent upper limits on the ultrahigh 
energy cosmic neutrino flux (cf. Fig.~\ref{cosmogenic_auger}).

The effective aperture for deeply penetrating showers has been parametrised 
in~\cite{Tu:2004ms}.
In short, we set $X_{\rm uno} (\theta) = X (\theta) - 1300~{\rm g /cm^2}$ and
$X_{\rm uno} \geq 1700~{\rm g /cm^2}$, in order to take into account the fact that on 
average a $10^{10}$~GeV shower traverses about $800~{\rm g /cm^2}$
before it develops to its maximum~\cite{Anchordoqui:1998nq}.
The detection efficiency $P (E_{\rm sh})$ reaches 100$\%$ at $E_{\rm sh} \approx 10^{10}$~GeV, where 
the aperture for electromagnetic showers induced by $\nu_e$ is found to be 
${\cal A}_{\rm eff} \sim 300~{\rm m^2\, sr}$~\cite{Yoshida:2001icrc}. The event rate at AGASA is given by
\begin{equation}
    \frac{d N}{d t} \equiv \int d E_{\rm sh}\,
    F (E_\nu)\, {\cal A}_{\rm eff} (E_\nu) \,\,,
\end{equation}
where
\begin{equation} 
{\cal A}_{\rm eff} (E_\nu) = A_p\, P (E_{\rm sh})\, {\rm att} (E_\nu)\, .
\label{charlyref}
\end{equation}
From Eq.~(\ref{charlyref}) one can reliably deduce the effective area $A_p.$
For $\nu_e$ electromagnetic showers we  
obtain $A^{\rm em}_p = 56~{\rm km^2},$ with $E_{\rm sh} = E_\nu.$ 
As there is no estimate of the aperture for hadronic showers available, to be
conservative hereafter we simply assume $A^{\rm had}_p = A^{\rm em}_p$.
At energies $< 10^{10}$~GeV, we interpolate between $P (E_{\rm sh}) = 0$ at $10^{8}$~GeV 
and $P (E_{\rm sh}) = 1$ at $10^{10}$~GeV.

\subsection{RICE}

The Radio Ice Cherenkov Experiment (RICE) at the South Pole
aims to detect electron neutrinos of energies $> 10^{6}$~GeV based
on the principle of ``radio coherence''. Namely,
the electrons produced in the $\nu_e$ charged current interactions 
($\nu_e N \rightarrow e^- X$) in the Antarctic ice-cap
initiate sub-showers which emit Cherenkov radiation over a wide range of 
electromagnetic frequencies. The RICE detector is sensitive only to the 
long-wavelength range.
Monte Carlo simulations show that RICE achieves an effective volume 
$V_{\rm eff} \sim 1~{\rm km}^3$ at 
$E_{\rm sh} \approx  10^{8.5}$~GeV and
approaches $V_{\rm eff} \sim 10~{\rm km}^3$ 
at higher energies \cite{Kravchenko:2002mm}.
Furthermore, since hadronic showers do not suffer significantly from the 
Landau-Pomeranchuck-Migdal (LPM) effect \cite{Alvarez-Muniz:1998px}, they 
can increase the detection efficiency of an under-ice radio Cherenkov detector
at energies in excess of $10^{6}$~GeV.
As no candidates for neutrino-induced events have been seen during 
data-taking in 1999, 
2000 and 2001 (3500 hrs lifetime), 95$\%$ CL
upper limits on the diffuse 
neutrino flux have been derived \cite{Kravchenko:2003tc}.

In order to calculate the expected event rates with Eq.~(\ref{rate_rice}),
we approximate the effective target volume of the RICE detector as follows.
Due to its location in ice (at 100 - 300 m depths), there are no hadronic backgrounds, so RICE utilises the whole zenith angular range 
$\theta = 0^{\circ} - 90^{\circ}$.  In our approach, the Monte Carlo 
effective volume for the electromagnetic and the hadronic cascades
given by the RICE Collaboration in~\cite{Kravchenko:2003tc} is approximated 
as $V_{\rm eff} (E) \approx \pi\, r^2 (E)\, z (E)$. 
The height of this cylinder is fixed to $z = 1$ km from the ice surface to 
below, since the most detection efficiency is in the upper km of the ice 
\cite{Kravchenko:2002mm}.
The zenith angle $\theta$ is measured from the centre of the detector, i.e. 
at a depth of 0.5 km.
With the knowledge of the detector geometry we are able to estimate the attenuation 
of the neutrino flux in the ice properly.

We have checked that, replacing our approximation for the RICE effective aperture 
in Eq.~(\ref{rate_rice}), 
we were able to reproduce the RICE upper limits on 
$\nu_e$ flux reported in \cite{Kravchenko:2002mm} to within $\approx 35\%$.
This may be further reduced (to $\approx 20\%$)
if one takes into account 
nuclear structure effects on the neutrino-nucleon charged current 
cross section~\cite{CastroPena:2000sx},
as was done in Ref.~\cite{Kravchenko:2002mm}.

Following the procedure introduced in Ref.~\cite{Anchordoqui:2002vb}, 
one can easily update the model-independent upper bounds on neutrino fluxes 
using the search result reported by the RICE 
Collaboration~\cite{Kravchenko:2003tc}. These new limits, which are shown in 
Fig.~\ref{cosmogenic_auger}, surpass the previous estimate derived by combining
Fly's Eye + AGASA exposures  by about one order of magnitude.

\subsection{PAO}

The Pierre Auger Observatory~\cite{Abraham}, which is actually comprised of two sub-observatories, 
will be the next large-scale neutrino detector in operation. The Southern site is currently operational
and in the process of growing to its final size of $A \simeq 3000$~km$^2$. Another site is 
planned for the Northern hemisphere. The PAO works in
a hybrid mode, and when complete, each site will contain 24 fluorescence detectors overlooking
a ground array of 1600 water Cherenkov  
detectors.
During clear, dark nights,  events
are simultaneously observed by
fluorescence light and particle detectors, allowing powerful reconstruction
and cross-calibration techniques. The first analyses of data from the PAO are 
currently underway~\cite{Anchordoqui:2004wb} and it is expected
that first results will be made public in the Summer of
2005 at the 29'th International Cosmic Ray Conference.

For standard neutrino interactions in the atmosphere, 
each site of PAO reaches about $1.3 \times 10^7$~kT sr of target mass
around $10^{10}$~GeV~\cite{Capelle:1998zz}.
The sensitivity of PAO for neutrino-induced (i.e. deeply penetrating) hadronic showers, 
defined as one event per year and per energy decade, 
is shown in Fig.~\ref{cosmogenic_auger} (dashed-dotted line).
An even greater acceptance~\cite{Bertou:2001vm} should be 
achievable for the case of Earth-skimming neutrinos which produce a 
$\tau$, that decays and generates a shower 
in the ground array.
The projected sensitivity for $\nu_{\tau}$ is also shown in Fig.~\ref{cosmogenic_auger}
(hatched area).
The prospect of tau neutrino detection by the PAO fluorescence detector 
has been also investigated~\cite{Aramo:2004pr}, though of course the acceptance is 
decreased because of the 
10\% duty cycle.

The rate of neutrino-induced events at the ground arrays of PAO can be 
calculated using Eq.~(\ref{eq:AGASA_check}). 
We estimate the effective aperture for neutrinos, 
\begin{equation}
\label{Auger_aperture}
    {\cal A}_{\rm eff} (E_{\nu}) \equiv 
    \frac{\sigma_{\nu N} (E_{\nu})}{\sigma^{\rm tot}_{\nu N} (E_{\nu})}\,
    {\rm att} (E_{\nu})\, P (E_{\rm sh})\, A_p (E_{\rm sh}) \, ,
\end{equation}
through a comparison with the geometric acceptance given 
in Ref.~\cite{Capelle:1998zz},
where ${\cal A}_{\rm eff} (E_{\nu}) \approx \sigma_{\nu N} (E_{\nu}) 
\times {\rm acceptance}~(E_{\rm sh})/m_p.$  The detection efficiency is found to be
\begin{equation}
P (E_{\rm sh}) = 0.654 \,\, \log_{10} [(E_{\rm sh}/1~{\rm GeV}) \times 10^9] - 10.9\,\,,
\end{equation}
for $E_{\rm sh} \leq 10^{8.9}$~GeV, and $P (E_{\rm sh}) = 1$ above this energy. The parameter $A_p$ in 
Eq.~(\ref{Auger_aperture})
is energy-dependent, because the PAO acceptance does not saturate  
in the entire energy range. 
Our selection criteria are
{\em i)} $75^\circ \leq \theta \leq 90^\circ$ for the zenith angle, and
{\em ii)} $X_{\rm max} \geq 2500~{\rm g / cm^2}$ for the shower 
maximum, which corresponds to requiring $X_{\rm uno} \geq 1700~{\rm g / cm^2}$ 
in our approach.
We consider showers with axis falling in the array.
The altitude of the PAO southern site (1200 m above sea level)
is also taken into account. 
We obtained $A^{\rm had}_p (E) \approx 
1.475~{\rm km}^2\, \left(E / 1~{\rm eV} \right)^{0.151}$
for hadronic showers above $\approx 10^{10}$~GeV, and
$A^{\rm EM}_p (E) \approx 7.037 \times 10^6~{\rm km}^2\,
\left(E / 1~{\rm eV} \right)^{-0.208}$ for electromagnetic showers.
As shown in Ref.~\cite{Capelle:1998zz}, the aperture for all showers 
(i.e. including showers with axis not going through the array)
is roughly 1.8 to 2.5 times larger.

\section{\label{flux} Ultrahigh Energy Cosmic Neutrino Fluxes}

The opacity of the CMB to ultrahigh energy protons 
propagating over cosmological distances guarantees a cosmogenic flux of neutrinos, 
originated via the decay of charged 
pions produced in the proton-photon interactions~\cite{Berezinsky:1969}. 
The intermediate state of the reaction $p + \gamma_{\rm CMB} \to N + \pi$ is dominated by 
the $\Delta^+$ resonance, because the $n$ decay length is smaller than the nucleon 
mean free path on the relic photons. Hence, there is roughly an equal number of $\pi^+$ 
and $\pi^0$. Gamma rays, 
produced via $\pi^0$ decay, subsequently cascade electromagnetically on the cosmic 
radiation fields 
through $e^+ e^-$ pair production followed by inverse Compton scattering. 
The net result is a pile up of $\gamma$ rays at GeV energies, 
just below the threshold for further pair production. 
On the other hand, each $\pi^+$ decays to 3 neutrinos and a positron. 
The $e^+$ readily loses its energy through synchrotron 
radiation in the cosmic magnetic fields. 
The neutrinos carry away about 3/4 of the $\pi^+$ energy, 
and therefore the energy in cosmogenic neutrinos is about 3/4 of the one 
produced in $\gamma$-rays.

The normalisation of the neutrino flux depends critically on the cosmological 
evolution of the cosmic ray
sources and on their proton injection spectra~\cite{Yoshida:pt}. 
Of course, the neutrino intensity  also depends on the homogeneity of
sources:  for example, semi-local objects, such as the Virgo
cluster~\cite{Hill:1985mk}, could contribute to the high
energy tail of the neutrino spectrum. 
Another source of uncertainty in the 
cosmogenic neutrino flux 
is the Galactic $\to$ extragalactic crossover energy of cosmic rays: 
while Fly's Eye 
data~\cite{Bird:1993yi} seem to favour a transition at $10^{10}$~GeV, a recent 
analysis of the HiRes 
data~\cite{Bergman:2004bk} points to a lower value $\sim 10^{9}$~GeV.
This translates into different proton luminosities at sources and consequently 
different predictions for the expected flux of neutrinos~\cite{Anchordoqui:2004eb}.

Very recently, some of us (FKRT) 
have performed an investigation of the actual size of the cosmogenic 
neutrino flux~\cite{Fodor:2003ph}.
The assumptions made therein were 
{\em i)} all observed cosmic ray events are due to protons, and
{\em ii)} their sources are isotropically distributed in the universe.
It was further assumed that all sources have identical injection spectra $J_p$, 
and that the redshift evolution of the source luminosity and of the source co-moving
number density can be parametrised by a simple power-law. 
Therefore, the co-moving emissivity of protons
injected with energy $E_i$ at a distance $r$ from Earth can be written as
\begin{equation}
\label{emissivity_distr}
   {\cal L}_p (r, E_i) = \rho_0\, \left( 1 + z (r) \right)^n\,
   \Theta (z - z_{\rm min})\, \Theta (z_{\rm max} - z)\, J_p (E_i)\, ,
\end{equation}
where the redshift $z$ and the distance $r$ are related by 
$d z = (1 + z)\, H (z)\, d r$. 
The Hubble expansion rate at a redshift $z$ is related to the present one $H_0$ through
$H^2 (z) = H^2_0\, \left[\Omega_M (1 + z)^3 + \Omega_{\Lambda} \right]$, where   
$\Omega_{M} = 0.3$ and $\Omega_{\Lambda} = 0.7$ were chosen.
The results turned out to be rather insensitive to the precise values of the 
cosmological parameters within their uncertainties. 
The redshift evolution index $n$ accounts for the evolution of source emissivities in
addition to the pure redshifting ($n = 0$).
The minimal and maximal redshift parameters $z_{\rm min}$ and $z_{\rm max}$ exclude the
existence of nearby and early time sources.
The values $z_{\rm min} = 0.012$ 
(corresponding to $r_{\rm min} = 50$ Mpc) and $z_{\rm max} = 2$ were chosen. 

The propagation of protons towards Earth can be well described~\cite{Bahcall:1999ap} by the
{\em propagation function} $P_{b|a} (E; E_i, r)$ (here $b = a = p$). 
It specifies the probability of detecting a particle of species $b$ above an energy $E$
on Earth due to one particle of species $a$ created at a distance $r$ with energy $E_i$.
To simulate the photohadronic processes of protons with the CMB photons, 
the SOPHIA Monte-Carlo program~\cite{Mucke:1999yb} was adopted. 
For $e^+ e^-$ pair production, the continuous energy loss approximation was used. 
With the help of the propagation function, 
the number of cosmic ray protons or cosmogenic neutrinos ($b = \nu$) 
arriving at Earth with energy $E$ per units of energy, area, 
time and solid angle can then be calculated by 
\begin{equation}
\label{flux_earth}
   F_b (E) \equiv \frac{d^4 N_b}{d E\, d A\, d t\, d \Omega}
   = \frac{1}{4 \pi} \int^{\infty}_0 d E_i \int^{\infty}_0 d r\, (-)
   \frac{\partial P_{b|p} \left(E; E_i, r \right)}{\partial E}\,
   {\cal L}_p (r, E_i)\, .
\end{equation}
This formula can also be easily generalised to arbitrary source emissivity. 

To determine the ``most probable'' cosmogenic neutrino flux $F_{\nu} (E)$
from the observed cosmic ray spectrum, 
an $E^{- \alpha}_i$ power-like injection spectrum for the protons was assumed
in Eq.~(\ref{emissivity_distr}) up to $E_{\rm max}$, the maximal energy attainable 
through astrophysical acceleration processes in a bottom-up scenario, i.e.,
\begin{equation}
\label{pow_inj}
   J_p (E_i) = J_0\, E^{- \alpha}_i\, \Theta (E_{\rm max} - E_i)\, .
\end{equation}
The predicted differential proton flux at Earth $F_p (E)$ 
was compared with the most recent observations by the AGASA \cite{Takeda:1998ps}
and the HiRes \cite{Abu-Zayyad:2002ta} experiments separately. 
A fitting procedure yielded the most probable values for the maximal proton
injection energy $E_{\rm max}$, the power-law index $\alpha$, and the redshift evolution
index $n$.
The factors $J_0$ and $\rho_0$ from 
Eq.~(\ref{emissivity_distr}) served to normalise the predicted proton flux. 

\begin{figure}[t!]
\vspace{-1.0cm}
\begin{center}
\includegraphics*[width=10.0cm,clip=]{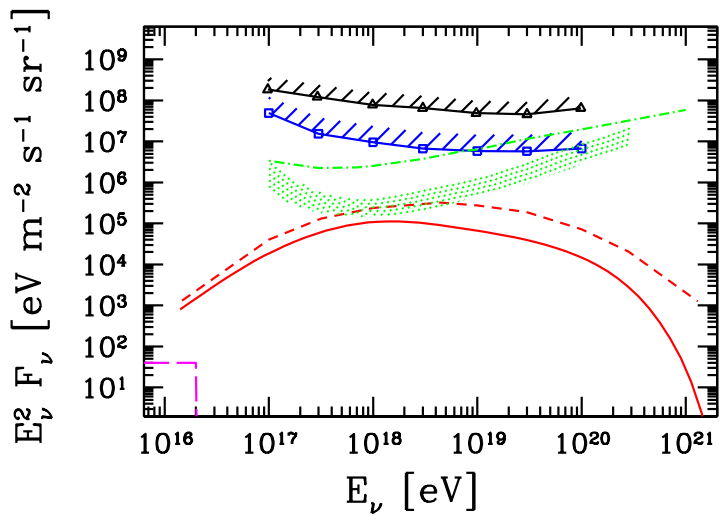}
\vspace{-1.0cm}
\caption[dum]
{Predictions for the cosmogenic neutrino flux per flavour. 
Solid line: flux from \cite{Fodor:2003ph} (FKRT) with the proton spectrum as
the best fit to the Akeno + AGASA cosmic ray data, corresponding to 
$\alpha = 2.57$, $n = 3.30$ and $E_{\rm max} = 10^{12.5}$~GeV. 
Dashed line: flux from \cite{Protheroe:1995ft} (PJ) assuming a maximum energy of 
$E_{\rm max} =  10^{12.5}$~GeV for the ultrahigh energy cosmic protons.
Long dashed line: flux of electron antineutrinos from neutron $\beta$--decay 
(parent heavy nucleus $^{56}$Fe, see Appendix).
Also shown are the 
differential upper limits on the ultrahigh energy neutrino flux per flavour.
Lower solid line with squares: upper limit by RICE, 
Eq.~(\ref{eq:sig_upp_modelind_RICE}). 
Upper solid line with triangles: the combined Fly's Eye + AGASA upper 
limit obtained in Ref.~\cite{Anchordoqui:2002vb}.
The dashed-dotted line and the hatched area are
the sensitivity (defined as one event per year and per energy decade)
of PAO for $\nu_e$ and for Earth-skimming 
$\nu_{\tau}$ \cite{Bertou:2001vm}, respectively.
}
\label{cosmogenic_auger}
\end{center}
\end{figure}

For each $E_{\rm max}$, the compatibility of various $(\alpha, n)$ pairs 
with the cosmic ray data in the energy range between $E_{-} = 10^{8.2}$~GeV 
and $E_{+} = 10^{11}$~GeV was checked.
This procedure gave the 
$2\sigma$ confidence regions in the $\alpha - n$ plane for each $E_{\rm max}$.
The best fit values are found to be $E_{\rm max} = 10^{12.5}$~GeV, 
$\alpha = 2.57$, $n = 3.30$ for AGASA, and $E_{\rm max} = 10^{12.5}$~GeV, 
$\alpha = 2.50$, $n = 3.80$, for HiRes.
The corresponding cosmogenic neutrino flux from fitting to AGASA data
is shown in Fig.~\ref{cosmogenic_auger} (solid line). The one from fitting 
to the HiRes data is smaller by roughly a factor 1.1 to 1.25, with larger $2\sigma$ 
uncertainties.

It should be noted that when performing  
the compatibility check of $(\alpha, n)$ with $E_{-} = 10^{9.5}$~GeV 
(and $E_{+} = 10^{11}$~GeV) for each $E_{\rm max}$, the redshift evolution 
index $n$ is no longer constrained in this case. 
To consider the possibility of a Galactic $\rightarrow$ 
extragalactic transition in agreement with the Fly's Eye data, in our analysis 
we adopt the cosmogenic neutrino flux estimates of
Protheroe and Johnson (PJ)~\cite{Protheroe:1995ft}.  
This analysis
incorporates the source cosmological evolution from
estimates~\cite{Rachen:1992pg} of the power per comoving volume
injected in protons by powerful radio galaxies.  
Here we use PJ 
$\nu_{\mu} + \bar{\nu}_{\mu}$ estimate with an injection spectrum with
$E_{\rm max} = 10^{12.5}~{\rm GeV}$. 
We stress that the PJ flux agrees with a most recent
estimate~\cite{Engel:2001hd} in the entire energy range, whereas the
spectrum obtained in earlier calculations~\cite{Yoshida:pt} is
somewhat narrower, probably as a result of different assumptions
regarding the propagation of protons. 

Cosmogenic antineutrinos can also be produced via decay of neutrons 
photo-dissociated from heavy nucleus primaries by the CMB. However, it turns 
out that antineutrinos from neutron $\beta$-decay contribute relatively 
little to the cosmogenic flux in the energy region of 
interest (see Appendix and Fig.~\ref{cosmogenic_auger}).

%%%%%%%%%%%%%%%%%%%%%%%%%%%%%%%%FIGURE%%%%%%%%%%%%%%%%%%%%%%%%%%
%%%%%%%%%%%%%%%%%%%%%%%%%%%%%%%%%%%%%%%%%%%%%%%%%%%%%%%%%%%%%%%%%

\section{\label{table}Upper Bounds on the Neutrino-nucleon Cross Section}

\begin{figure}[t!]
\vspace{-1.cm}
\begin{center}
\includegraphics*[width=10.0cm,clip=]{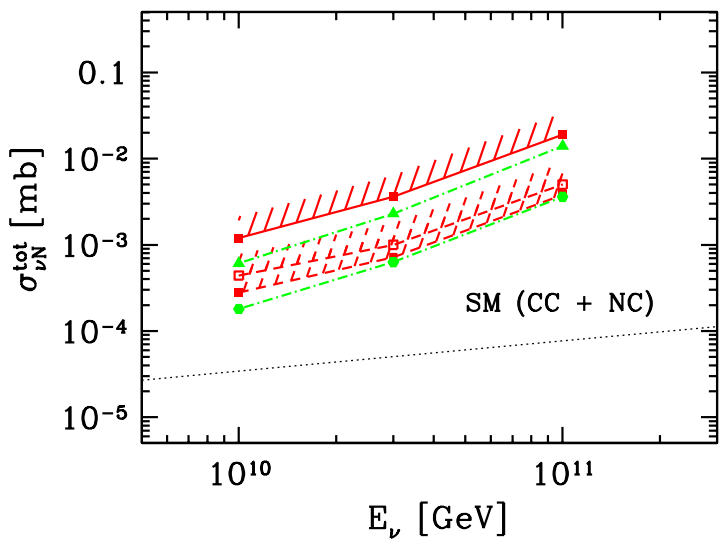}
\vspace{-1.3cm}
\caption[dum]
{Model-independent upper bounds on the neutrino-nucleon inelastic cross 
section derived from the
RICE Collaboration search results \cite{Yoshida:2001icrc}, by exploiting the
cosmogenic neutrino flux estimates by FKRT (solid line) and
PJ (dashed line joining solid squares).  
To give an idea of the scaling with the inelasticity parameter, 
the dashed line joining the open squares (PJ) indicates the upper bound for $\left< y \right> =0.5$. 
The dashed-dotted lines indicate the sensitivity (95\% CL, for  
$\sigma_{\nu N}^{\rm tot} <4$~mb) 
of PAO in 10 yr of operation assuming zero events observed above SM background
(circles PJ, triangles FKRT). 
For comparison, also shown is the SM total (charged current and 
neutral current) neutrino-nucleon inelastic cross 
section~\cite{Gandhi:1998ri}.}
\label{sig_upp_RICE}
\end{center}
\end{figure}

Upon their arrival at Earth,
ultrahigh energy cosmic neutrinos interact with nucleons  
at centre-of-mass energies $\sqrt{s}$ approaching several hundreds of TeV.
They probe the energy regime far beyond the reach of terrestrial colliders.
In the SM, ultrahigh energy neutrinos scatter deep-inelastically on nucleons. 
The double differential DIS-cross section $d^2 \sigma / dx dQ^2$
can be calculated with the help of the 
structure functions, where $x$ and $- Q^2$ are the 
Bjorken variable and the invariant momentum transfer, respectively.
The kinematic region in $(Q^2, x)$ probed thereby is 
$Q^2 \sim m^2_W \simeq 6.4 \times 10^3$~GeV$^2$
and $x = Q^2 / (y s) \approx 
1.7 \times 10^{-7} / (E_{\nu} / 10^{11}~{\rm GeV})$,
not accessible by HERA or other collider experiments.
It is therefore necessary to reliably extrapolate the nucleon structure 
functions to the region of high $Q^2$ and very small $x$ values. 
The cross sections for the $\nu N$ charged and neutral current interactions
have been estimated by several approaches
\cite{Gandhi:1998ri,Gluck:1998js,Kwiecinski:1998yf,Kutak:2003bd}.
All predict a power-like growth behaviour with energy (cf. Fig.~\ref{sig_upp_RICE} for
one of the predictions \cite{Gandhi:1998ri}). 
This reflects simply the rapid increase of parton densities towoards small Bjorken $x$,
as predicted by perturbative QCD and confirmed by HERA data.
Uncertainties due to different extrapolation approaches 
are about 30$\%$, and a factor of two at $E_{\nu} = 10^{12}$ GeV
if the gluon saturation effect is taken into account \cite{Kutak:2003bd}
(see e.g. \cite{Tu:2004ms} for a brief discussion on this).

Beyond the SM, the uncertainties in the neutrino interaction are conceivably larger.
In many extensions of the SM, neutrinos can interact with nucleons via additional 
channels, the rates for which exceed the SM one largely \cite{Domokos:1998ry}.
If neutrinos become comparably strongly interacting as protons 
at an energy around $10^{10.6}$~GeV, 
ultrahigh energy cosmic neutrinos could have already manifest themselves 
as the observed cosmic ray events beyond the Greisen--Zatsepin--Kuzmin (GZK) 
cutoff. The idea of using this ``strongly interacting'' neutrino 
scenario~\cite{Berezinsky:1969} 
to solve the GZK puzzle is supported by the observation that the cosmogenic neutrino
flux well matches the observed ultrahigh energy cosmic ray spectrum above 
the GZK energy $E_{\rm GZK} \approx 10^{10.9}$~GeV (see e.g. FKRT 
in Ref.~\cite{Domokos:1998ry} for a statistical analysis of this scenario).

The neutrino-nucleon inelastic cross section at ultrahigh energies
is constrained by several considerations. As $s$ goes to infinity,
the unitarity bound \cite{Froissart:1961ux}
limits the cross section to grow at most as
$\sigma^{\rm tot} \leq C \cdot \ln^2 s.$ 
However, without the knowledge of the constant 
$C$ this bound cannot be of practical use.
On the other hand, the neutrino-nucleon cross section at high energies is related to 
the low-energy neutrino-nucleon elastic amplitude through 
dispersion relations\cite{Goldberg:1998pv}. 
Laboratory neutrino-nucleon fixed-target scattering experiments at relatively low
energies can therefore indirectly observe or constrain anomalous enhancements of
$\sigma_{\nu N}$ at high energies.  
The non-observation of ultrahigh energy neutrino-induced events reported by several
experiments such as Fly's Eye, AGASA and RICE imposes  upper bounds on the
neutrino-nucleon inelastic cross section as well. They depend, however, sensitively 
on the neutrino flux.  

In what follows we derive upper limits on the neutrino-nucleon inelastic 
cross section
by exloiting the predictions for the cosmogenic neutrino flux discussed 
in Sec.~\ref{flux} and the formulae presented in Sec.~\ref{formulae}.
First we note that Eq.~(\ref{eq:AGASA_check}) and  Eq.~(\ref{rate_rice})
can be used to constrain new physics models which predict
an enhancement of the SM neutrino-nucleon cross section above some energy 
threshold.
The ``strongly interacting neutrino'' scenarios are also subject to this constraint.
If the jump of the neutrino-nucleon cross section is not strictly a step function, 
neutrinos with energies in the intermediate range, where the cross section is
$\sim 0.1~{\mu}$b - $0.5$ mb, can always initiate showers deep in the 
atmosphere~\cite{Anchordoqui:2000uh}.

Now we derive   
model-independent upper bounds on $\sigma_{\nu N}^{\rm tot} (E_{\nu})$ from AGASA's 
search result on quasi-horizontal air showers \cite{Inoue:cn,Yoshida:2001icrc}.
From Eq.~(\ref{eq:AGASA_check}), we demand \cite{Anchordoqui:2002vb,Tu:2004ms}
\begin{equation}
\label{eq:sig_upp_modelind_AGASA}
   A_p\, t\, \left< P (E_{\rm sh}) \right>\, 
   \left<E_{\nu}\, F_{\nu} (E_{\nu}) \right>\,
   \left< {\rm att} (E_{\nu}) \right> < 3.5 / \Delta\, ,
\end{equation}
in a sufficiently small interval $\Delta$, where a single power law 
\begin{equation}
   P (E_{\rm sh})\, F_{\nu} (E)\, {\rm att} (E) \propto E^{\gamma}\, , 
\end{equation}
is valid.
The choice $\Delta = 1$, corresponding to one $e$-folding of energy, is reasonable.
In Table~\ref{tb:sig_limit_AGASA} we present our results by exploiting the 
cosmogenic neutrino fluxes discussed in the previous section. 
We have assumed that the neutrino-nucleon interactions do not distinguish between
different flavours, and that the total neutrino energy
goes into visible shower energy, i.e. $E_{\rm sh} = E_{\nu}$.

We found that these bounds are applicable only for 
$\sigma^{\rm tot}_{\nu N} \lesssim 0.5~{\rm mb}$. If neutrinos 
interact more strongly, they would induce air showers high in 
the atmosphere,
thus avoid the bounds derived from the non-observation of quasi-horizontal
air showers by ground arrays.
They would instead contribute to the vertical showers. 
Therefore, for energies $E_{\nu} \gtrsim 10^{11}$ GeV, the AGASA search results on
the horizontal showers cannot constrain the neutrino-nucleon cross section, 
if the neutrino fluxes are at the level as the cosmogenic ones we exploited.

%%%%%%%%%%%%%%%%%%%table%%%%%%%%%%%%%%%%%%%%%%%%%%%%%%%%%%%%%%%%%%%%%%
\begin{table}[t!]
\begin{center}
\begin{tabular}{|c|c|c|}
\hline
  $\,\,\,\,$ $E_{\nu}$ [GeV] $\,\,\,\,$ & $\,\,\,\,$ FKRT \cite{Fodor:2003ph} $
\,\,\,\,$
  & $\,\,\,\,$ PJ \cite{Protheroe:1995ft} $\,\,\,\,$ \\
\hline
\hline
  $10^{10}$ & $1.3 \times 10^{-1}$ & $2.4 \times 10^{-2}$ \\
\hline
  $10^{10.5}$ & no & $1.4 \times 10^{-1}$  \\
\hline
  $10^{11}$ & no & no \\
\hline
\end{tabular}
\caption[Upper bounds on the neutrino-nucleon cross section]
{Upper bound on the neutrino-nucleon cross section (in [mb])
derived from the AGASA Collaboration search results \cite{Yoshida:2001icrc}.}
\label{tb:sig_limit_AGASA}
\end{center}
\end{table}
%%%%%%%%%%%%%%%%%%%%%%%%%%%%%%%%%%%%%%%%%%%%%%%%%%%%%%%%%%%%%%%%%%%%%%%

Next we derive upper bounds on the neutrino-nucleon cross section from RICE's search
results \cite{Kravchenko:2002mm,Kravchenko:2003tc} using Eq.~(\ref{rate_rice}).
We have checked again that, in the interval $\Delta = 1$, 
the single power law approximation $F_{\nu} (E)\, \sigma_{\nu N}^{\rm tot} (E)\, 
V (E)\, {\rm atten} (E) \propto E^{\alpha}$ is valid, where we define 
\begin{equation}
   {\rm atten} (E) \equiv \int^{90^{\circ}}_{0^{\circ}} d \theta\,
   2 \pi \sin \theta\, e^{- \frac{x (\theta)\, \sigma_{\nu N}^{\rm tot} (E)}{m_p}}\, .
\end{equation}
We demand 
\begin{equation}
\label{eq:sig_upp_modelind_RICE}
   \frac{t^\prime\, \rho_{\rm ice}}{m_p}\, 
   \left<E_{\nu}\, F_{\nu} (E_{\nu}) \right>\, 
   \left< \sigma_{\nu N}^{\rm tot} (E_{\nu}) \right>\, 
   \left< {\rm atten} (E_{\nu}) \right>\,
   \left< V (E_{\rm sh}) \right> < 3.09 / \Delta \, .
\end{equation}
We use the data taken 
in 1999, 2000 and 2001 \cite{Kravchenko:2003tc}, where
the total lifetime is $t^\prime = 3500$ hrs.
The upper bounds on the neutrino-nucleon cross section 
from the RICE Collaboration search results are listed in Table~\ref{tb:sig_limit_rice} and given in 
Fig.~\ref{sig_upp_RICE}.
For comparison, one of the predictions for the SM total cross section \cite{Gandhi:1998ri}
is also shown.
We found that the bounds derived from RICE's search result are applicable for
$\sigma^{\rm tot}_{\nu N} \lesssim 1~{\rm mb}$. 
%150205ht corrected.

%150205ht corrected.
%%%%%%%%%%%%%%%%%%%table%%%%%%%%%%%%%%%%%%%%%%%%%%%%%%%%%%%%%%%%%%%%%%
\begin{table}[t!]
\begin{center}
\begin{tabular}{|c|c|c|}
\hline
$\,\,\,\,$ $E_{\nu}$ [GeV] $\,\,\,\,$ & $\,\,\,\,$ FKRT \cite{Fodor:2003ph} $\,\,\,\,$
  & $\,\,\,\,$ PJ \cite{Protheroe:1995ft} $\,\,\,\,$\\
\hline
\hline
  $10^{10}$ & $1.2 \times 10^{-3}$ & $2.8 \times 10^{-4}$   \\
\hline
  $10^{10.5}$ & $3.6 \times 10^{-3}$ & $7.2 \times 10^{-4}$   \\
\hline
   $10^{11}$ & $1.9 \times 10^{-2}$ & $3.8 \times 10^{-3}$  \\
\hline
\end{tabular}
\caption[dum]
{Upper bound on the neutrino-nucleon cross section (in [mb])
derived from the RICE Collaboration search 
results~\cite{Kravchenko:2002mm,Kravchenko:2003tc}.
}
\label{tb:sig_limit_rice}
\end{center}
\end{table}
%%%%%%%%%%%%%%%%%%%%%%%%%%%%%%%%%%%%%%%%%%%%%%%%%%%%%%%%%%%%%%%%%%%%%%%

%%%%%%%%%%%%%%%%%%%%%%%%%%%%%%%%FIGURE%%%%%%%%%%%%%%%%%%%%%%%%%%
%%%%%%%%%%%%%%%%%%%%%%%%%%%%%%%%%%%%%%%%%%%%%%%%%%%%%%%%%%%%%%%%%

\section{\label{PAO} Sensitivity of PAO to Anomalous Neutrino Interactions}

%%%%%%%%%%%%%%%%%%%table%%%%%%%%%%%%%%%%%%%%%%%%%%%%%%%%%%%%%%%%%%%%%%
\begin{table}[t!]
\begin{center}
\begin{tabular}{|c|c|c|}
\hline
$\,\,\,\,$ Event rates $\,\,\,\,$ & $\,\,\,\,$ FKRT \cite{Fodor:2003ph}
 $\,\,\,\,$ & $\,\,\,\,$ PJ \cite{Protheroe:1995ft} $\,\,\,\,$\\
\hline
\hline
  $\nu_{\mu} + \nu_e + \nu_\tau$ & $0.06 - 0.09$ & $0.16 - 0.25$  \\
\hline
  $\nu_{\tau}$ & $0.19 - 0.54$  & $0.56 - 1.59$  \\
\hline
\end{tabular}
\caption[dum]
{Yearly neutrino event rates expected at the Southern site of PAO, 
for the cosmogenic neutrino flux estimates by 
PJ and by FKRT.
Numbers in the first line are 
for hadronic showers falling within the array
and all (hadronic) showers, respectively, induced by all neutrino (+ antineutrino) flavours
in the range $10^{9}~{\rm GeV} < E_\nu < 10^{11}~{\rm GeV}.$
Numbers in the second line are for Earth-skimming tau neutrinos in the range
$10^{9}~{\rm GeV} \leq E_{\nu} \leq 10^{11}~{\rm GeV}$, obtained by 
assuming strong energy loss due to deep
inelastic scattering (DIS), and no DIS-loss \cite{Bertou:2001vm}.
}
\label{tb:rate_auger}
\end{center}
\end{table}
%%%%%%%%%%%%%%%%%%%%%%%%%%%%%%%%%%%%%%%%%%%%%%%%%%%%%%%%%%%%%%%%%%%%%%%

In this section we estimate the potential of PAO to probe physics beyond the 
SM. Given the prospects for fairly high statistics, detailed analyses of high 
energy neutrino interactions are in principle possible. In particular, if no 
enhancement of deeply developing showers is observed, PAO will be able to set 
stringent bounds on the neutrino-nucleon inelastic cross section. 

In order to estimate 
these bounds, we again adopt the cosmogenic neutrino fluxes shown in Fig.~\ref{cosmogenic_auger}
and assume that only SM sources of deeply developing showers are observed. 
Note that in contrast to AGASA, SM neutrino interactions lead to observable 
rates at PAO. In Table~\ref{tb:rate_auger} 
we list the 
yearly SM neutrino event rates to be expected at
the PAO,
for the cosmogenic neutrino fluxes we reviewed in Sec.~\ref{flux}.
To obtain these event rates we used   
the SM cross section estimate given in Fig.~\ref{sig_upp_RICE}~\cite{Gandhi:1998ri}. 

Before proceeding, we verify whether hadronic showers may be a significant background 
in our analysis. The number of cosmic ray showers expected to be detected by PAO 
in the angular bin $\theta \in (75^\circ, 90^\circ)$ is given by
\begin{equation}
\frac{dN_p}{dt} = A\, \int_{75^\circ}^{90^\circ} \cos \theta \,\,d\Omega \,
\int_{E_1}^{E_2} \,\, P(E,\Omega) \,\, E^3\,\, F_p^{\rm obs} (E)\,\, \frac{dE}{E^3} \,\,,
\label{pao1}
\end{equation}
where  $F_p^{\rm obs}(E)$ is the incoming flux of cosmic rays and $E_1$ and $E_2$ are the minimum and 
maximum energy under considerations.
For $E_1 = 10^{10}$~GeV, the PAO detection efficiency $P(E,\Omega)$ reaches 100\%, and so   
Eq.~(\ref{pao1}) can be rewritten as
\begin{equation}
\frac{dN_p}{dt} \approx A \,\,\Delta \Omega \,\,\,\langle E^3 F_p^{\rm obs}(E) \rangle \,
\,\frac{1}{2\,E_1^2} \,\,.
\label{pao2}
\end{equation}
Now replacing in Eq.~(\ref{pao2}) the observed isotropic flux in this region, 
$\langle E^3 F_p^{\rm obs}(E)\rangle = 
10^{24.5 \pm 0.2}$~eV$^2$ m$^{-2}$ s$^{-1}$ sr$^{-1}$~\cite{Anchordoqui:2002hs}, we obtain 
$dN_p/dt \approx 317$~yr$^{-1}.$ A detailed background event estimate requires a convolution with the 
detector resolution, and it is beyond the scope of this paper. However, the $\langle X_{\rm max} \rangle$ 
distribution obtained~\cite{Gaisser:1993ix} through Monte Carlo simulations indicates 
that the probability  a proton-induced 
shower with $10^{10}~{\rm GeV} < E < 
10^{11}~{\rm GeV}$ leads to $X_{\rm max } > 2500$~g/cm$^2$ is $< 10^{-4},$ hence hereafter we neglect the 
hadronic background in our analysis.  

With this in mind, as a very conservative estimate of background events, we consider the 
expected SM neutrino 
showers with $ 10^{9}~{\rm GeV} < E_\nu < 10^{11}~{\rm GeV}$, given in Table~\ref{tb:rate_auger}. 
In Table~\ref{tb:pao} we list the 
sensitivity of PAO (95\% CL corresponding to 3.54, 4.24 
events for FKRT and PJ, respectively~\cite{Feldman:1997qc}) 
for anomalous neutrino cross sections after 
5 yr of operation. In Fig.~\ref{sig_upp_RICE} the 10 yr prospects ($N = 3.96, 5.08$, for FKRT and PJ, 
respectively) to improve these bounds are shown. Note that in 12 yr of running, the SM background 
of deeply developing showers at the Southern site will correspond to 3.09 events. In such a case, 
another technique to bound the rise of the cross section can be used. Namely, by separately binning 
events which 
arrive at very small angles to the horizontal and
comparing event rates of deeply developing showers and Earth-skimmers, 
the neutrino-nucleon cross section can be inferred~\cite{Anchordoqui:2001cg,Kusenko:2001gj}. 
This is because the flux of up-going $\tau$'s per unit surface area produced by Earth-skimming neutrinos 
is {\it inversely proportional} to $\sigma_{\nu N}^{\rm tot},$ whereas the rate of deeply developing showers 
due to neutrino interactions in the atmosphere is {\it proportional} to  
$\sigma_{\nu N}^{\rm tot}.$ Therefore, we conclude that the full observatory (Northern 
and Southern sites) in 6~yr of running will reach the sensitivity to probe cross sections at the level of 
SM predictions.

%%%%%%%%%%%%%%%%%%%table%%%%%%%%%%%%%%%%%%%%%%%%%%%%%%%%%%%%%%%%%%%%%%
\begin{table}[t!]
\begin{center}
\begin{tabular}{|c|c|c|}
\hline
$\,\,\,\,$ $E_{\nu}$ [GeV] $\,\,\,\,$ & $\,\,\,\,$ FKRT \cite{Fodor:2003ph} $\,\,\,\,$
  & $\,\,\,\,$ PJ \cite{Protheroe:1995ft} $\,\,\,\,$\\
\hline
\hline
  $10^{10}$ & $1.0 \times10^{-3} - 1.9 \times 10^{-3}$ & $3.1 \times 10^{-4} - 5.6 \times 10^{-4}$ \\
\hline
  $10^{10.5}$ & $4.2 \times 10^{-3} - 8.7 \times 10^{-3}$ &
$1.1 \times 10^{-3} - 2.1 \times 10^{-3}$\\
\hline
   $10^{11}$ & $2.7 \times 10^{-2} - 7.8 \times 10^{-2}$ & $6.2 \times 10^{-3} - 1.5 \times 10^{-2}$\\
\hline
\end{tabular}
\caption[dum]
{Sensitivity (95\% CL) of PAO for the neutrino-nucleon cross section 
(in [mb]) derived assuming non-observation of deeply developiong 
showers above SM background in 5 yr of operation.
}
\label{tb:pao}
\end{center}
\end{table}
%%%%%%%%%%%%%%%%%%%%%%%%%%%%%%%%%%%%%%%%%%%%%%%%%%%%%%%%%%%%%%%%%%%%%%%

\section{\label{conclusions} Conclusions}

The search for ultrahigh energy cosmic neutrinos is entering an exciting era.
Recently, a wide array of projects have been initiated to detect neutrinos by 
searching for low-altitude quasi-horizontal showers, cascades in the Antarctic ice-cap, 
and radio emission from neutrino-induced showers. Some of these experiments, like  
PAO and RICE, have been taken data, and others, like the km$^3$ IceCube 
telescope, are under construction.

Although such a high energy cosmic neutrino flux has not been observed to date, it 
has long been known that there should be a guaranteed cosmogenic flux resulting 
from the interactions of cosmic protons with the CMB. Therefore, the non-observation 
of deeply developing showers or cascades in ice reported by neutrino-detection-experiments 
in conjunction with these cosmogenic neutrinos can be used to constrain the behaviour 
of the neutrino-nucleon inelastic cross section in the energy regime far beyond the reach 
of terrestrial colliders. 

In this work we derived model-independent upper bounds on the 
neutrino-nucleon inelastic cross section from existing search results 
reported by the AGASA and the RICE collaborations.
The bounds apply for neutrino-nucleon cross sections smaller than 0.5~mb 
(AGASA), or 1~mb (RICE). Interestingly, the search result reported by the 
RICE Collaboration improves the upper bounds derived from the non-observation 
of deeply developing showers at AGASA by more than one order of 
magnitude. Therefore, in the presence of additional neutrino-emitting-sources, 
there is only  little room for new physics contributions to the inelastic 
cross section. 

We have also estimated the sensitivity of PAO to neutrino interactions. This hybrid 
detector will facilitate powerful air shower reconstruction methods and control of 
the systematics errors which have plagued cosmic ray experiments to date. Through a 
comparison  of deeply developing showers and Earth-skimming events, 
in 6 yr of operation the observatory will be able to probe neutrino interactions at 
the level of SM predictions, providing a final verdict on the 
rise of the neutrino-nucleon inelastic cross section at ultrahigh energies.

\section*{Acknowledgments}

We thank Markus Ahlers,
Veniamin Berezinsky, Jonathan Feng, Haim Goldberg, 
Francis Halzen, Karl-Heinz Kampert, Al Shapere, Anna Stasto, and Tom Weiler for some valuable discussions. 
This work was partially 
supported by the US National Science Foundation (grant No. PHY-0140407) and 
the Hungarian Science Foundation (grants No. OTKA-T34980/37615/46925/TS44839).

\section*{Appendix}

If ultrahigh energy cosmic rays are heavy nuclei, the relic photons can excite the 
giant dipole resonance at nucleus energies $E \gsim 10^{11}$~GeV, and thus there should be 
accompanying photo-dissociated free neutrons, themselves a source of $\beta$-decay 
antineutrinos. 
The decay mean free path of a neutron is 
$c \Gamma_n \overline \tau_n = (E_n/10^{11}$~GeV)~Mpc, the lifetime being boosted from its 
rest frame value $\overline \tau_n = 886$~s to its lab value via $\Gamma_n = E_n/m_n$. 
Compared to cosmic distances  $\gsim 100$~Mpc, the decay of even the boosted neutron 
may be taken 
as nearly instantaneous, and thus all free neutrons are themselves a source 
of $\beta$-decay cosmogenic antineutrinos. The neutron emissivity 
${\cal L}_n (E_n),$  defined
as the mean number of neutrons emitted per co-moving volume per unit time per unit energy 
as measured at the source can be estimated as follows.
Neutrons with energies above $10^{9.3}$~GeV have parent iron nuclei with
$\Gamma  > \Gamma_0 \approx 2 \times 10^{9}$ which are almost completely disintegrated
in distances of less than 100~Mpc~\cite{Epele:1998ia}.\footnote{Because of the position of 
$^{56}$Fe in the binding energy curve, it is generally considered to be a significant end 
product of stellar evolution, and indeed heavy mass nuclei are found to be be much rare in 
the cosmic radiation. Thus, here we adopt iron as the pivot nucleus species, i.e., $A =56.$} 
Thus, it is reasonable to define a characteristic time  $\tau_{_{\Gamma}}$ given by the
moment at which the number of nucleons is reduced to $1/e$ of its 
initial value $A$, and presume  the nucleus, emitted at distance $d$ from the Earth,  
is a travelling source that at  $D \approx (d - c \tau_{_{\Gamma}})$
disintegrates into $A$ nucleons all at once~\cite{Anchordoqui:1997rn}. Then, 
the number of neutrons with energy 
$E_n = E_A/A$ 
can be approximated by the product of 1/2 the number of nucleons generated per nucleus 
and the number of nuclei emitted, i.e., ${\cal L}_n (E_n) = N {\cal L}_A$, 
where $N = A - Z$ is the mean neutron 
number of the source nucleus. Now, to obtain an estimate of the diffuse antineutrino 
flux one needs to integrate over the 
population of nucleus-emitting-sources out to the horizon~\cite{Tom}
\begin{eqnarray}
F_{\overline \nu} (\Enu) \equiv  \frac{d\Fnu}{d\Enu}(\Enu) & = & \frac{1}{4\pi\,H_0}\, 
   \int dE_n\,{\cal L}_n (E_n) \,\,
   \left[1-\exp\left(-\frac{D\,m_n}{E_n\,\overline \tau_n}\right) \right] \,\, 
   \int_0^Q d\epsilon_{\overline \nu}\,\frac{dP}{d\epsilon_{\overline \nu}}
   (\epsilon_{\overline \nu})  \nonumber \\
   &  & \times \int_{-1}^1 \frac{d\cos \overline \theta_{\overline\nu}}{2} \; 
   \delta\left[E_{\overline \nu}-E_n\,\epsilon_{\overline \nu}\,(1+\cos 
  \overline \theta_{\overline \nu}) /m_n\right] \,,
\label{nuflux}
\end{eqnarray}
where the
$r^2$ in the volume integral 
is compensated by the usual $1/r^2$ fall-off of flux per source.
Here, $H_0$ is the Hubble constant, $E_{\overline \nu}$ and $E_n$ are the antineutrino and
neutron energies in the lab, $\overline \theta_{\overline \nu}$ is
the antineutrino angle with respect to the direction of the
neutron  momentum in the neutron rest-frame, and $\epsilon_{\overline \nu}$ is
the antineutrino energy in the neutron rest-frame.  The last three variables are not observed
by a laboratory neutrino-detector, and so are integrated over.
The observable $\Enu$ is held fixed.
The delta-function relates the neutrino energy in the lab to the
three integration variables.
The parameters appearing in Eq.~(\ref{nuflux}) are the
neutron mass and rest-frame lifetime ($m_n$ and $\tbar$). Finally, $dP/d\Enubar$ is the
normalised probability that the
decaying neutron produces a $\overline \nu$ with
energy $\Enubar$  in the neutron rest-frame. Note that the maximum $\overline \nu$ energy 
in the neutron 
rest frame is very nearly  $Q \equiv m_n - m_p - m_e = 0.71$~MeV.
Integration of Eq.~(\ref{nuflux})  can be 
easily accomplished, especially when two good approximations are 
applied~\cite{Anchordoqui:2003vc}.
The first approximation is to think of the $\beta$--decay as a $1 \to 2 $ process of 
$\delta m_N \to e^- + \overline \nu,$ in which the neutrino is produced 
monoenergetically in the rest frame, with $\epsilon_{\overline \nu} = \epsilon_0 
\simeq \delta m_N (1  - m_e^2/ \delta^2 m_N)/2 \simeq 0.55$~MeV, where 
$\delta m_N \simeq 1.30$~MeV 
is the neutron-proton mass difference. In the lab,
the ratio of the maximum $\overline \nu$ energy to the neutron energy  
is $2 \epsilon_0/m_n \sim 10^{-3},$ 
and so the boosted $\overline \nu$ has a spectrum with 
$E_{\overline\nu} \in (0, 10^{-3} \, E_n).$ 
 The second approximation is to replace the neutron decay probability 
$1 - e^{-Dm_n/E_n \overline \tau_n}$
with a step function $\Theta (E_n^{\rm max} - E_n)$ at some energy 
$E_n^{\rm max} \sim {\cal O}(D \, m_n/\overline{\tau}_n) = 
(D/10~{\rm Mpc}) \times 10^{12}$~GeV. 
Combining these two approximations we obtain
\begin{equation}
\frac{d\Fnu}{d\Enu}(\Enu) = \frac{m_n}{8\,\pi\, \epsilon_0 \,H_0} 
\int_{E_A^{\rm min}}^{E_A^{\rm max}} \frac{dE_A}{E_A/A}\,\, {\cal L}_A (E_A)\,\,,
\label{S1}
\end{equation}
where $E_A^{\rm min} \equiv {\rm max} \{ E_{A,\Gamma_0}, \frac{A\, 
m_n E_{\overline \nu}}{2 \epsilon_0} \},$ and $E_A^{\rm max}$ is the energy 
cutoff at the nucleus-emitting-source $\ll A (D/10~{\rm Mpc}) \times 10^{12}$~GeV.
For ${\cal L}_A \propto E_A^{-\alpha},$ integration of 
Eq.~(\ref{S1}) leads to 
\begin{equation}
    \frac{d\Fnu}{d\Enu}(\Enu) \approx 10^6 \left(\frac{E_{A,\Gamma_0}}
    {E_A^{\rm max}}\right)^\alpha 
    \left[\left(\frac{E^{\rm max}_{\overline \nu}}{E_{\overline \nu}}
    \right)^\alpha -1 \right] \, \left. 
    \frac{d{\cal F}_A}{dE_A} \right|_{\Gamma_0}\,\,,
\label{znz}
\end{equation}
where $E_{\overline\nu} \gsim  10^{6.3} \,(56/A)$~GeV and
\begin{equation}
E_{\overline \nu}^{\rm max} = \frac{2 \epsilon_0}{m_n}\,\, 
\frac{E_A^{\rm max}}{A} \sim  10^{7.3}\,\, \left(\frac{56}{A}\right)\,\,
\left(\frac{E_A^{\rm max}}{10^{12}~{\rm GeV}}\right) \,\, {\rm GeV}\,.
\end{equation}
The sub-PeV antineutrino spectrum is flat as all the free neutrons have sufficient energy
$E_n \gsim E_{\Gamma_0}/A,$ to contribute equally to all the $\overline \nu$ energy 
bins below $10^{6}$~GeV. Taking $\alpha = 2$ as a reasonable example, and inputting the 
observational value $\left. E^2_{A,\Gamma_0} \,\,d{\cal F}_A^{\rm obs} /dE_A \right|_{\Gamma_0} 
\approx 10^5$~eV m$^{-2}$ s$^{-1}$ sr$^{-1}$~\cite{Anchordoqui:2002hs} Eq.~(\ref{znz}) 
becomes~\cite{Tom}
\begin{equation}
    E_{\overline \nu}^2 \frac{d\Fnu}{d\Enu}(\Enu) \approx 4 
    \times 10^{1}\,\, \left(\frac{56}{A}\right)\,\,{\rm eV}\, {\rm m}^{-2}\,\, 
    {\rm s}^{-1}\,\, {\rm sr}^{-1} \,. 
\end{equation}
Note that the $\beta$-decay process gives initial antineutrino flavour ratios $1 : 0 : 0$ 
and Earthly ratios nearly $3:1:1.$ Compared to full-blown Monte Carlo 
simulations~\cite{Hooper:2004jc}, 
this paper--and--pencil calculation underestimates the flux by about 30\%. 
Of course the situation described above represents the most extreme case, 
in which all cosmic rays at the end of the spectrum are heavy nuclei. 
A more realistic guess would be that the composition at the 
end of the spectrum is mixed.

\end{document}